\documentclass[twocolumn,secnumarabic,amssymb, nobibnotes, aps]{revtex4-1}
\usepackage{graphicx}
\usepackage{dcolumn}
\usepackage{bm}
\usepackage{ucs}
\usepackage{float}
\usepackage{amsmath}

\begin{document}

\title{Scaling of sub-gap excitations in a  
superconductor-semiconductor nanowire quantum dot}
\author{Eduardo J. H. Lee$^{1}$}
\thanks{Current address: Condensed Matter Physics Center (IFIMAC), Universidad Aut\'{o}noma de Madrid, 28049 Madrid, Spain.}
\email{eduardo.lee@uam.es}
\author{Xiaocheng Jiang$^{2}$}
\author{Rok \v{Z}itko$^{3}$}
\author{Ram\'{o}n Aguado$^{4}$}
\author{Charles M. Lieber$^{2}$}
\author{Silvano De Franceschi$^{1}$}
\email{silvano.defranceschi@cea.fr}

\affiliation{$^{1}$SPSMS, CEA-INAC/UJF-Grenoble 1, 17 rue des Martyrs, 38054 Grenoble Cedex 9, France}
\affiliation{$^{2}$Harvard University, Department of Chemistry and Chemical Biology, Cambridge, MA, 02138, USA}
\affiliation{$^{3}$ Jo\v{z}ef Stefan Institute, Jamova 39, SI-1000, Ljubljana, Slovenia}
\affiliation{$^{4}$Instituto de Ciencia de Materiales de Madrid (ICMM), Consejo Superior de Investigaciones Cient\'{i}ficas (CSIC), Sor Juana In\'{e}s de la Cruz 3, 28049 Madrid, Spain}

\date{\today}%

\begin{abstract}
A quantum dot coupled to a superconducting contact provides a tunable artificial analogue of a magnetic atom in a superconductor, a paradigmatic quantum impurity problem. We realize such a system with an InAs semiconductor nanowire contacted by an Al-based superconducting electrode. We use an additional normal-type contact as weakly coupled tunnel probe to perform tunneling spectroscopy measurements of the elementary sub-gap excitations, known as Andreev bound states or Yu-Shiba-Rusinov states. We demonstrate that the energy of these states, $\zeta$, scales with the ratio between the Kondo temperature, $T_K$, and the superconducting gap, $\Delta$. $\zeta$ vanishes for $T_K/\Delta \approx 0.6$, denoting a quantum phase transition between spin singlet and doublet ground states. By further leveraging the gate control over the quantum dot parameters, we determine the singlet-doublet phase boundary in the stability diagram of the system.  Our experimental results show remarkable quantitative agreement with numerical renormalization group calculations.
\end{abstract}

\maketitle

A magnetic impurity coupled to a metal reservoir, as described by the Anderson impurity model, provides the theoretical basis for important phenomena in condensed matter, such as the Kondo effect or strongly correlated materials. Recently, the specific case wherein the metal reservoir is a superconductor (S) has attracted considerable interest, largely due to its relevance in the context of non-trivial superconducting states. 
Indeed, theoretical proposals suggest that the bound states known as Andreev levels or Yu-Shiba-Rusinov (YSR) states that appear in this limit \cite{Yu, Shiba, Rusinov}, are precursors of a 1D topological superconductor with zero-energy Majorana edge modes \cite{Nadj-Perge, KlinovajaRKKY, Franz2013, Braunecker, Pientka, Fulga2013, SauDasSarma, ZhangNori}. However, in spite of its importance, quantitative experimental studies of the S-coupled Anderson impurity remain scarce \cite{Buizert, Wernsdorfer12, Pillet2013, NygardABS}. In particular, the scaling of Andreev levels with respect to the relevant physical parameters (e.g., the tunnel coupling between S and the impurity, $\Gamma_S$) has not yet been addressed. 
Here, we present a joint experimental-theoretical work aimed at filling this void. We exploit the versatility of semiconductor quantum dots (QDs), which effectively behave as a quantum impurity, to investigate the scaling of Andreev levels in a direct manner, by tunneling spectroscopy. Our quantitative analysis is further supported by numerical renormalization group (NRG) calculations performed without fitting parameters, which show remarkable agreement with the measured data.

The ground state of the S-coupled Anderson impurity is defined in  
a competition involving the superconducting proximity effect, Coulomb blockade and Kondo correlations.  
There are two possibilities: a magnetic doublet, enforced by strong Coulomb interactions, and a spin singlet, favored by strong coupling to S. 
Transitions between the ground state and the first excited state of the system, i.e. between a doublet and a singlet state, or vice-versa, are manifested as a sub-gap Andreev level of energy $\zeta$, where the latter is equivalent to the excitation energy. 
Remarkably, the theory predicts that $\zeta$ scales with $\Gamma_S$, and that a quantum phase transition (QPT) between singlet and doublet ground states takes place when $\zeta$ changes sign (signaled by the crossing of Andreev levels at zero energy) \cite{Glazman, Rozhkov99, Vecino03, Simon, Hewson04, Hewson07, Choi}. 
In this work, we employ tunneling spectroscopy to study the Andreev levels associated with a QD formed in hybrid superconductor-semiconductor nanowire structures. With the aid of a dual-gate device geometry, we are able to continuously tune $\Gamma_S$ while probing the same QD charge state. 
In this way, we demonstrate full electrical control over Andreev levels as well as over the singlet-doublet QPT. 
By further studying the evolution of Andreev levels in the parameter space, we obtain an experimental phase diagram of the system, and verify that the tuning of Andreev levels is consistent with the predicted scaling with the dimensionless ratio between the Kondo temperature and the superconducting gap, $T_K/\Delta$. 
We note that while a similar tuning of the QPT has been indirectly studied in the supercurrent behavior of QD-based Josephson junctions \cite{Wernsdorfer12}, here we provide the first spectroscopical demonstration, which is also fully supported by numerically exact NRG simulations. Finally, we point out that the herein discussed formation of QDs in hybrid nanowire devices, and the sensitivity of device parameters to the local electrostatic environment are relevant effects to be considered in experiments aimed at the detection of Majorana modes.

\begin{figure}
\includegraphics[width=86mm]{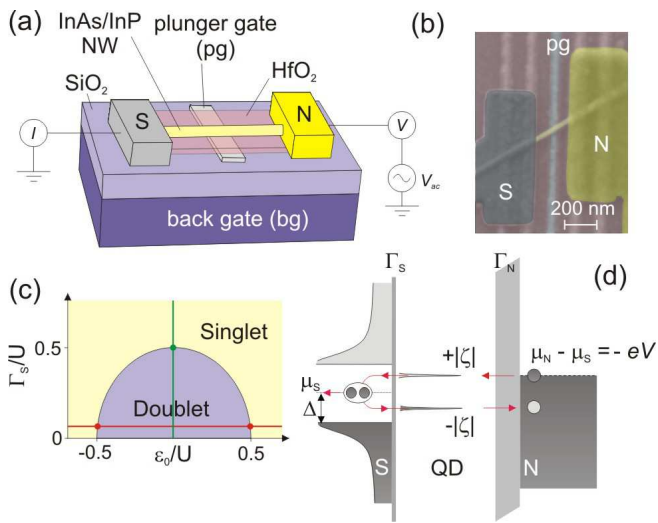}
\caption{(a) Schematics of the studied dual-gate N-QD-S devices. The nanowires are contacted by N and S leads comprised of Ti/Au and Ti/Al bilayers, respectively. A Ti/Au thin strip covered by HfO$_2$ dielectrics acts as a local plunger gate (pg), whereas the degenerately-doped substrate is employed as a global back gate (bg). (b) False color scanning electron micrograph of a typical device (top-view). (c) Qualitative phase diagram of the QD-S system in the wide-gap limit ($\Delta \rightarrow \infty$). The horizontal (vertical) line  underscores QPTs between the singlet and doublet states (circles) that occur upon varying the QD level position (QD-S coupling). (d) Schematics of the Andreev level spectroscopy transport cycle. Current is measured across the N-QD-S device when the chemical potential of the tunnel probe ($\mu_N$) is aligned with an Andreev level at energies $\pm|\zeta|$. Transport occurs via Andreev reflection, whereby an injected electron (hole) is reflected back to N as a hole (electron), forming (breaking) a Cooper pair in S.}
\end{figure}

A scheme of the device geometry adopted in this study is shown in Fig.~1a, where N represents a normal metal tunnel probe weakly-coupled to the QD. InAs/InP core/shell NWs grown by thermal evaporation \cite{Xiaocheng} were randomly dispersed onto highly-doped Si/SiO$_2$ substrates (300 nm-thick oxide) containing pre-patterned local bottom gate arrays. 
Individual wires sitting on top of local gates were identified by scanning electron microscopy. Source and drain contacts were subsequently defined by standard e-beam lithography techniques, followed by metal deposition and lift-off. The finalized devices contained a single local gate, later used as a plunger gate (pg), 
between the N (2.5 nm Ti/45 nm Au) and S (2.5 nm Ti/45 nm Al) contacts. In the experiment, the dual-gate action is achieved by employing the degenerately-doped substrate as a global back gate (bg). A representative scanning electron micrograph of a typical device is shown in Fig.~1b. We note that in our devices, single QDs formed spontaneously in the NW segments located between the electrical leads. Tunneling spectroscopy was performed by first ensuring that the coupling asymmetry strongly favored the S lead (i.e., $\Gamma_S \gg \Gamma_N$). In this regime, the differential conductance, $dI/dV$, measured as a function of the source-drain bias, $V$, reflects the density of states of the QD-S system. Measurements were carried out using conventional lock-in techniques ($V_{ac}$ = 5 $\mu$V) in a dilution fridge operating at a base temperature of 13 mK.

The ground state of the QD-S system is defined by an interplay of the relevant energy scales: $\Gamma_S$, the QD level position, $\epsilon_0$, the on-site charging energy, $U$, and $\Delta$. An intuitive picture of the underlying competition can be gained by considering the limiting cases. In the weak coupling limit, when Coulomb blockade is the dominant effect ($\Gamma_S \ll U$), a one by one charge filling of the dot is enforced, thereby stabilizing the doublet state. The singlet state, by its turn, has two limiting characters. 
In the wide superconducting gap limit ($\Delta \rightarrow \infty$), 
strong QD-S couplings ($\Gamma_S \gg U$) favor Cooper pairs to occupy the dot, leading to a BCS-like singlet ground state. By contrast, for $\Gamma_S > \Delta$, 
the ground state is a Kondo-like singlet. A precise boundary between the different singlet states is however not well-defined \cite{Yamada}.

Phase diagrams depicting the stability of the possible ground states can be theoretically calculated by considering the above energy scales \cite{Rozhkov99, Vecino03, Simon, Hewson04, Hewson07,Choi, Yamada}. Fig.~1c shows a qualitative example for illustrative purposes. This simple diagram already captures two important features. First, it shows that for a constant $\Gamma_S$, which is 
the typical situation in an experiment, QPTs between the singlet and doublet states occur by sweeping the QD level position (red line). For a fixed $\epsilon_0$, on the other hand, the QPT can be driven by tuning $\Gamma_S$ (green line).

\begin{figure}
\includegraphics[width=86mm]{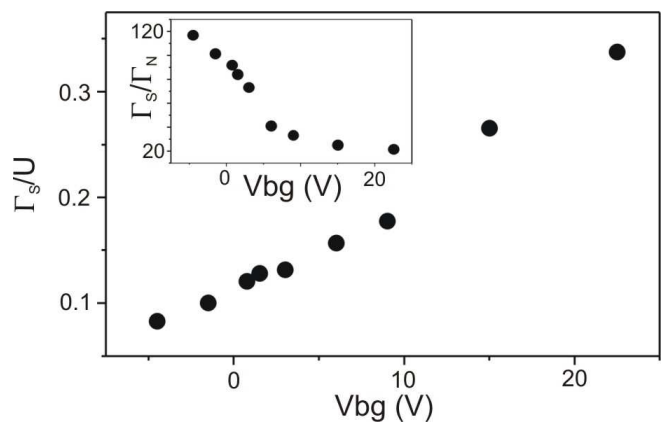}
\caption{Effect of $V_{bg}$ on the QD-S tunnel coupling, $\Gamma_S/U$ (main panel), and coupling asymmetry, $\Gamma_S/\Gamma_N$ (inset). The device parameters were obtained by fitting the normal state data (see text for more details). The plots demonstrate a continuous back gate-induced tuning of $\Gamma_S/U$. N behaves as a tunnel probe irrespective of $V_{bg}$, even if $\Gamma_S/\Gamma_N$ is also affected by the back gating.}
\end{figure}

\begin{figure*}
\includegraphics[width=170mm]{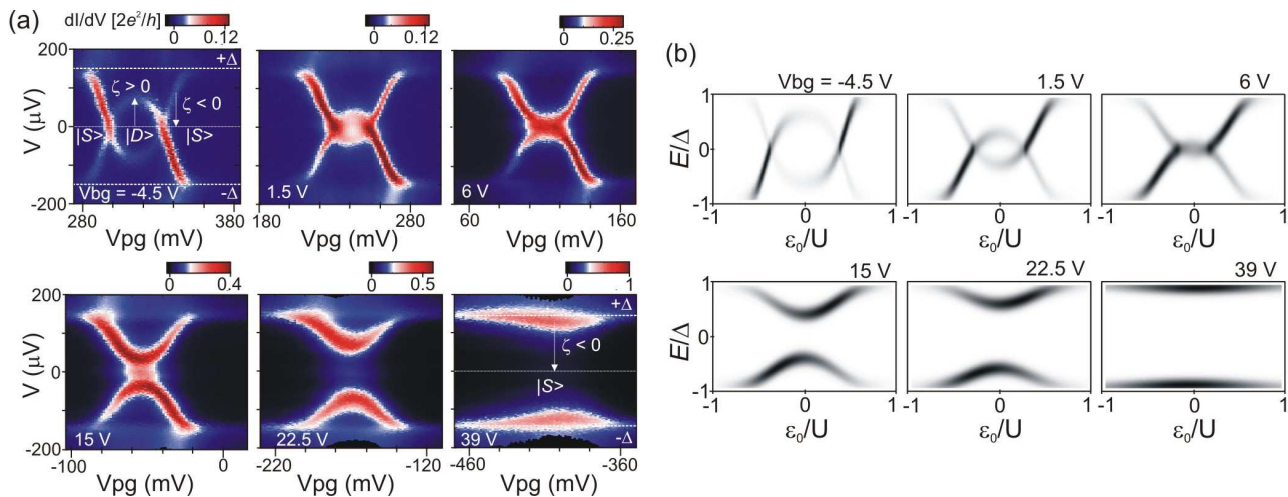}
\vspace{3mm}
\caption{(a)Series of $dI/dV$ vs. ($V,V_{pg}$) plots depicting the impact of back gating on the energy of Andreev levels. $V_{bg}$ increases from -4.5 V to 39 V. The horizontal lines in the top left panel highlight the positions of the superconducting gap, $eV = \Delta$, and the Fermi level, $eV = 0$. $|S>$ and $|D>$ refer to the singlet and doublet ground states, respectively. We adopt the convention that the Andreev level energy, $\zeta$, is positive (negative) for a doublet (singlet) ground state. The doublet ground state region is gradually suppressed for increasing $V_{bg}$, suggesting a QPT induced by the electrical tuning of $\Gamma_S$. (b) Density of states spectra of the proximity-coupled QD calculated by NRG using $U$, $\Gamma_S$ and $\Delta$ extracted from the experimental data.}
\end{figure*}

Transitions between the ground state and the excited state are detected by tunneling spectroscopy \cite{Pillet, Pillet2013, Mason, Lee2012, Deacon2010, Kanai, Kumar, Lee2014} as pairs of Andreev level resonances symmetrically positioned around the Fermi level, 
as illustrated in Fig.~1d. Here, we adopt the convention that $\zeta > 0$ for a doublet ground state.  
When the chemical potential of the tunnel probe, $\mu_N$, is aligned with an Andreev level, an electron (or a hole) tunnels into it, 
which changes the fermion parity of the proximitized dot. This is followed by an Andreev reflection process, in which a second electron (hole) enters the dot forming (breaking) a Cooper pair in S, while reflecting a hole (electron) back to N. As a result, the QD-S system relaxes back to its initial state, and measurable current is detected through the device. 

We start the experiment by suppressing the superconductivity in the S lead with  a small out-of-plane magnetic field ($B_\bot$ = 30 mT). 
Normal state charge stability diagrams (not shown) were taken by measuring $dI/dV(V)$ as a function of the plunger gate voltage, $V_{pg}$. Odd occupancy states were selected by identifying Coulomb diamonds displaying 
a zero-bias Kondo ridge.  
In the following, the discussion will be focused on 
a specific device. 
Data corresponding to a second device can be found in the Supplemental Material.

We then evaluated the impact of the back gate on the parameters of the studied devices (Fig.~2). We have found that fits to the linear conductance $dI/dV(V=0, V_{pg})$ provided a significantly more accurate estimate of $\Gamma_S$ when compared to evaluating it from $T_K$ values obtained from the width of the Kondo resonance as performed in ref. \cite{Wernsdorfer12} (see Supplemental Material).
In addition to $\Gamma_S$, we have also reliably extracted the values for the coupling asymmetry, ($\Gamma_S/\Gamma_N$), and $U$ as a function of the back gate voltage, $V_{bg}$. The charging energy is only weakly affected by the back gate, decreasing from $\approx$ 2.5 meV to $\approx$ 1.98 meV as $V_{bg}$ is swept from -4.5 V to 22.5 V. Most importantly, a sizable and continuous gate-induced tuning of $\Gamma_S$ is demonstrated in the main panel of Fig.~2. 
Notably, the $V_{bg}$-dependence of the coupling asymmetry reveals that $\Gamma_N$ is also affected by the back gate (inset).
Nevertheless, the conditions $\Gamma_N/\Gamma_S \ll 1$ and $\Gamma_N \ll U$ are always fulfilled, ensuring the role of a weakly coupled tunnel probe for the N contact.

We now turn to the $dI/dV(V,V_{pg})$ measurements acquired in the superconducting state ($B_\bot = 0$). For even occupancy, the low-bias conductance is drastically suppressed due to the absence of quasiparticles 
within $\Delta$; the gap appears to be relatively hard \footnote{Given the device geometry, the QD couples most likely to the portion of the semiconductor nanowire covered by the S contact. In this hypothesis, {$\Delta$} corresponds to the induced superconducting gap.}.
The onset of transport is heralded by the $dI/dV$ peaks at $eV \approx \pm \Delta \approx \pm$ 150 $\mu$V corresponding to the onset of quasiparticle tunneling above the superconducting gap edge.
The estimated $\Delta$ = 150 $\mu$eV is consistent with previously reported values for similar devices \cite{Lee2012, Chang}.

Odd-occupancy states display a much richer sub-gap structure. Fig. 3a shows a series of plots corresponding to the same odd charge state but taken at different $V_{bg}$, hence different $\Gamma_S/U$ values. To gain a better understanding of their meaning, we start by discussing the top left panel in greater detail.
The most remarkable features are pronounced sub-gap $dI/dV$ peaks that show a striking $V_{pg}$ dependence. These peaks can be ascribed to Andreev levels appearing at energies $eV = \pm |\zeta|$. Their gate modulation reveals a marked sensitivity of $\zeta$ with respect to the QD level position. Of particular interest are the two points where the Andreev levels cross at zero bias. They represent degeneracies between the singlet and doublet states where the QPTs take place. An intuitive picture of the above interpretation can be grasped by recalling that the $V_{pg}$ range covered in the measurement is qualitatively equivalent to that represented by the horizontal line in the phase diagram in Fig. 1c. 
Specifically, as $V_{pg}$ is swept to more positive voltages from the left, the ground state changes twice upon crossing the phase boundaries. 
Importantly, the observation of two crossings is consistent with a measurement taken at a relatively weak QD-S coupling.

The following panels in Fig.~3a reveal a clear trend for increasing $\Gamma_S/U$. This corresponds to an upward shift of the horizontal line in the phase diagram of Fig. 1c. At first, the zero-bias crossing points move closer together, signaling that the doublet region shrinks. By further increasing $V_{bg}$ to 6 V, the two crossings merge into a single point approximately positioned at the center of the Coulomb diamond. For even higher $\Gamma_S/U$, the Andreev levels no longer cross, suggesting that the singlet becomes the ground state throughout the entire $V_{pg}$ range.
We notice that an unexpected feature emerges in the strong coupling, singlet regime. It consists of a zero-bias $dI/dV$ peak, which is clearly visible for $V_{bg}$ = 15V and persists at higher $V_{bg}$ where it gets overshadowed by the increasing magnitude of the Andreev resonances (the latter is due to the increasing
values of $\Gamma_N$ and $\Gamma_S$, and the color scale has been adjusted accordingly
). A similar zero-bias feature was observed in ref. \cite{Chang} and a possible explanation in terms of a Kondo-type anomaly was suggested \cite{ZitkoPRB}. This interpretation may hold also is the present case. Since the available data do not allow us to go beyond this speculative level, we shall not discuss this observation any further.

Altogether, the behavior of the sub-gap levels shown in Fig.~3a demonstrates a QPT driven by the electrical tuning of $\Gamma_S/U$, which is corroborated by the $V_{bg}$-dependence of device parameters. To provide further support to our interpretation, we have simulated the density of states spectra of the QD-S system by feeding the experimentally measured device parameters to NRG calculations. The numerical results, presented in Fig. 3b, show remarkably good agreement with the experimental data.

\begin{figure}
\includegraphics[width=86mm]{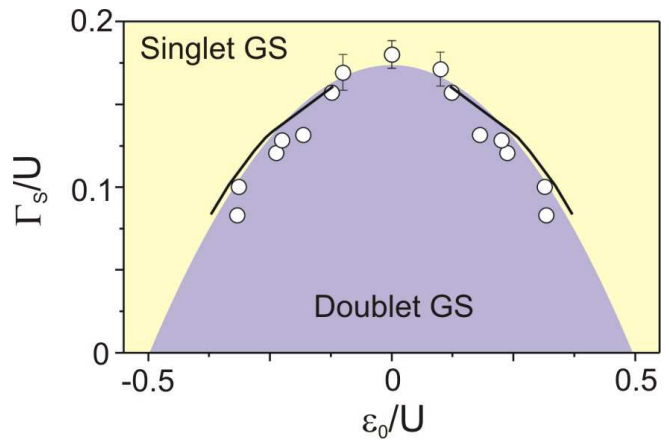}
\caption{Experimental phase diagram of the QD-S system. The parameter space is composed of the QD-S tunnel coupling, $\Gamma_S$, and the QD level position, $\epsilon_0$, scaled by $U$. The open dots represent the phase boundaries extracted from the experimental data. The three data points located around $\epsilon_0/U = 0$ were estimated from the $\zeta=0$ intercept in $\zeta(\Gamma_S/U)$ traces. The error bars are the associated errors in these fits. The remaining points were obtained directly from the gate-dependent Andreev level tunnel spectra. The solid lines represent phase boundaries simulated by NRG. The colored limits of the singlet and doublet states are guides to the eye.}
\end{figure}

As a subsequent step, we gathered the information contained in Figs. 2 and 3 in the form of an experimental phase diagram (Fig. 4). Two different methodologies were used to estimate the experimental phase boundaries (open circles). The most straightforward method relied on directly tracking the position of Andreev level crossings in $dI/dV(V,V_{pg})$ plots taken at fixed $V_{bg}$, to extract the $\epsilon_0/U$ coordinates of the boundaries. These were later associated with the corresponding $\Gamma_S/U$ coordinates obtained from Fig. 2. However, owing to the finite width of the Andreev levels, this task became increasingly difficult as the crossing points moved closer together. To circumvent this issue, the phase boundaries around the particle-hole symmetry point were estimated from the $\zeta=0$ intercept in $\zeta(\Gamma_S/U)$ plots, which were obtained from measurements at constant $\epsilon_0$. The resulting experimental diagram shows a remarkable quantitative agreement with the phase boundaries obtained from the NRG calculations (solid lines).

\begin{figure}
\includegraphics[width=86mm]{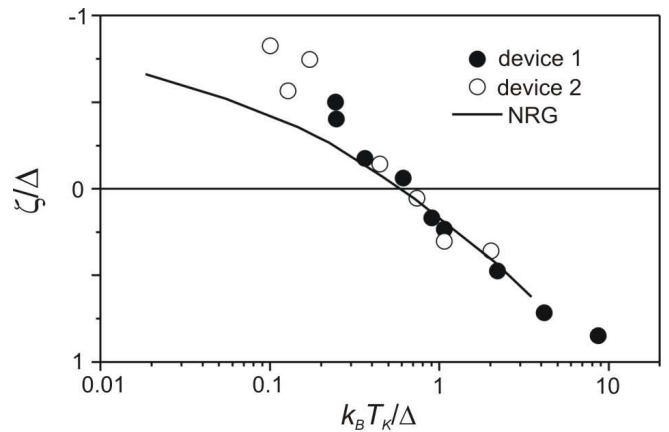}
\caption{$T_K/\Delta$ scaling of the Andreev level energy, $\zeta/\Delta$. The QPT occurs at $\zeta = 0$. Two datasets are shown: one corresponding to the device discussed in the main text (device 1, closed dots) and another of the device shown in the Supplementary Material (device 2, open dots). The solid line is the scaling curve calculated by the NRG.}
\end{figure}

Finally, we study the scaling of the Andreev levels 
with respect to $T_K/\Delta$. For this analysis, we used $T_K$ values estimated from the half-width of the normal state Kondo resonances measured at the center of the Coulomb diamonds ($\epsilon_0$ = 0). The Andreev level energy at the same position, $\zeta(\epsilon_0 = 0)$, is plotted against $T_K/\Delta$ in Fig. 5, which includes data from a second device (device 2, presented in more detail in the Supplemental Material). Interestingly, both data sets display nearly identical scaling which, for $T_K/\Delta \gtrsim 0.3$, also shows an excellent agreement with the NRG calculations. From the intersection of the data with $\zeta = 0$, we estimate that the QPT occurs 
at $T_K/\Delta \approx 0.6$. 
This value agrees with those reported in the literature, taking into account the differences in the definition of $T_K$ used in various works.
By contrast, the discrepancy between the experiment and the theory for low $T_K/\Delta$ is attributed to an overestimation of $T_K$ in the weak coupling limit.
Indeed, by taking into account $B_\bot =$ 30 mT and $g$-factor $\sim$ 6, as measured in a similar device \cite{Lee2014}, the Zeeman energy is expected to be of the order of 10 $\mu$eV. While this energy scale is negligible compared to the width of the Kondo resonance in the strong coupling limit, it becomes comparable to $T_K$ on the weak coupling side (leftmost points in Fig. 5), leading to an appreciable broadening of the Kondo resonance.

Our herein reported findings of electrically-tunable Andreev levels, in combination with a previous demonstration of their spin polarization \cite{Lee2014}, constitute important milestones towards pursuing proposals of engineering topological superconductors from arrays of proximity-coupled QDs.

\section*{Acknowledgements}

We acknowledge financial support from the Spanish Ministry of Economy and Competitiveness through grant Nos. FIS2012-33521, FIS2015-64654-P (MINECO/FEDER). R\v{Z} acknowledges the support of the Slovenian Research Agency (ARRS) through grants P1-0044 and J1-7259.

\bibliography{lee2016}

\end{document}